\documentstyle[psfig]{elsart}

\def\gray{$\gamma$-ray}
\def\grays{$\gamma$-rays}

\def\apj{Astrophys. J.}

\def\mqg{M_{\rm QG}}
\def\mpl{M_{\rm Planck}}

\def\epr{e-print}

\begin{document}

\begin{frontmatter}

\title{Constraints on Lorentz Invariance Violating Quantum Gravity and Large 
Extra Dimensions Models using High Energy Gamma Ray Observations}

\author{F.W. Stecker}

\address{Laboratory for High Energy Astrophysics, NASA Goddard Space Flight 
Center, Greenbelt, MD 20771, USA}

\begin{abstract}

Observations of the multi-TeV spectra of the nearby BL objects Mkn 421 and Mkn 
501 exhibit the high energy cutoffs predicted to be the result of 
intergalactic annihilation interactions, primarily with infrared photons 
having a flux level as determined by various astronomical observations. 
After correction for this absorption effect, the 
derived intrinsic spectra of these multi-TeV sources can be explained within 
the framework of simple synchrotron self-Compton emission
models. Stecker and Glashow have shown that the existence of such 
annihilations {\it via} electron-positron pair production interactions up to 
an energy of 20 TeV puts strong constraints on Lorentz invariance 
violation. Such constraints have important implications for 
Lorentz invariance violating (LIV) quantum gravity models as well as LIV 
models involving large extra dimensions. 
We also discuss the implications of observations of high energy \grays\ from 
the Crab Nebula on constraining quantum gravity models.

\end{abstract}


\end{frontmatter}

\section{Introduction}

Absorption of high energy \grays\ from extragalactic sources occurs {\it via} 
interactions of these photons with low energy photons of intergalactic 
radiation \cite{st92}, \cite{ss98}. These low energy photons are produced by 
stellar radiation and the reemission of such radiation by interstellar dust in 
galaxies. These photons then leave the galaxies in which they were produced, 
escaping into intergalactic space. The absorption is expected to produced
by the annihilation of high energy \grays\ with low energy photons into  
electron-positron pairs in intergalactic space. Stecker and Glashow \cite{sg01}
have shown that the very existence of such interactions puts quantitative
constraints on Lorentz invariance violation. We show here that high
energy \gray\ obsevations can also be used to place significant constraints
on proposed extra dimensions and quantum gravity scenarios.

\section{Observations of the Blazars Mkn 501 and Mkn 421}

The highest energy extragalactic \gray\ sources in the known universe are
the active galaxies called `blazars', objects that emit jets of relativistic 
plasma aimed directly at us.
Those blazars, known as X-ray selected BL Lac objects (XBLs), or
alternatively as high frequency BL Lac objects (HBLs), are expected to emit
photons in the multi-TeV energy range, but  only the nearest ones
are expected to be observable at TeV energies, the others being hidden by 
intergalactic absorption \cite{st96}.

Extragalactic photons with the highest energies yet observed originated 
in a powerful flare coming from the giant elliptical active galaxy known as 
Mkn 501 \cite{ah99}. Its spectrum is most easily understood and interpreted
as manifesting the high energy absorption to be expected from \gray\ 
annihilation by extragalactic pair production interactions.
The analyses of de Jager and Stecker \cite{ds02}
and Konopelko {\it et al.} \cite{ko03} indicate the presence of the
absorption effect in agreement with that predicted by calculating the expected 
energy dependent opacity inferred from the background light spectral energy
distributions derived in reference \cite{ma01} and reference \cite{ds02}.
This absorption is the result of electron-positron 
pair production by interactions of the multi-TeV \grays\ from Mkn 501 
primarily with intergalactic infrared photons.
Intrinsic absorption by pair production interactions within \gray\ sources 
such as Mkn 421 and Mkn 501 is expected to be negligible because such 
giant elliptical galaxies contain little dust to emit infrared radiation 
and because BL Lac objects have little gas (and therefore most likely 
little dust) in their nuclear regions. It also appears that \gray\ 
emission in blazars takes place at superluminal knots in 
their jets, downstream of the radio cores of these active galaxies and 
therefore downstream of any putative accretion disks \cite{jo01}.

The spectrum of Mkn 501 in the flaring phase extends 
to an energy of at least 24 TeV \cite{ah99}. The calculations of de Jager
and Stecker \cite{ds02} predict that intergalactic absorption should strongly 
suprress the spectra of these sources at multi-TeV energies. The intrinsic 
spectrum of Mkn 501 was derived by by these authors, correcting
for the opacity calculated for $z = 0.03$ as a function of energy, based on 
models of MS01, extended into the optical and UV range. 

Konopelko, {\it et al.} \cite{ko03} have reexmained the spectra of both Mkn 
421 and Mkn 501 corrected for absorption and have found that they fit  
synchrotron self-Compton (SSC) emission models from the radio to TeV 
\gray\ range. Both the X-ray and intrinsic TeV spectra of Mkn 421 peak at
lower energies than those of the flaring spectrum of Mkn 501, 
which can be understood if electrons were accelerated to higher
energies in the April 1997 flare of Mkn 501 than in Mkn 421.
The intrinsic spectrum of Mkn 501 with
the absorption effect removed actually peaks at multi-TeV energies, rather than
falling off in this energy range as observed \cite{ah99}, \cite{kr99}. Thus, 
it appears that the dropoff in 
the observed \gray\ spectrum of Mkn 501 above $\sim$ 5 TeV is a direct 
consequence of intergalactic absorption. We will therefore interpret the Mkn 
501 data as evidence for intergalactic absorption with no indication of 
Lorentz invariance violation (see next section) up to a photon energy of 
$\sim\,$20~TeV. This conclusion can be substantiated by observations of other
extragalactic \gray\ sources at differing redshifts \cite{ss98}.

\section{High Energy Consequences of Breaking of Lorentz Invariance} 

It has been suggested that Lorentz invariance (LI) may be only an approximate 
symmetry of nature \cite{sa72} - \cite{co98}. A simple 
and self-consistent framework for analyzing possible departures from exact LI 
was suggested by Coleman and Glashow \cite{co99}, who assume LI to be broken
perturbatively in the context of conventional quantum field theory. Small
Lorentz noninvariant terms are introduced that are renormalizable, being
of mass dimension no greater than four and having dimensionless coupling
constants in the kinetic part of the Lagrangian. These terms are also chosen
to be gauge invariant under $SU(3)\times SU(2)\times U(1)$ 
(``the almost standard model''). It is further 
assumed that the Lagrangian is rotationally invariant in a preferred frame
which is presumed to be  the rest frame of the cosmic microwave background.

Consequent observable manifestations of LI breaking 
can be described quite simply in terms of different
maximal attainable velocities of different particle species as
measured in the preferred frame. This is because the small LI violating
terms modify the free-field propagators so that the maximum velocities  
of various particles are not equal to $c$. 
It follows that if LI is violated the 
maximum attainable velocity of an electron need not equal the velocity of 
light {\it in vacuo\/}, {\it i.e.,}  $c_e \ne c_\gamma$. The physical
consequences of this violation of LI depend on the sign of the difference 
between $c_e$ and $c_{\gamma}$ \cite{co99}. Following the discussion of 
Stecker and Glashow \cite{sg01}, we define

\begin{equation}
c_{e} \equiv c_{\gamma}(1 +  \delta) ~ , ~ ~~~0< |\delta| \ll 1\;,  
\end{equation}

\noindent
and consider the two cases of positive and negative values of $\delta$
separately.\footnote{Note that the parmeter $\delta$ defined here differs
from that defined in reference \cite{co99} of $\delta_{\gamma e} \equiv
c_{\gamma}^2 - c_{e}
^2 \simeq 2(c_{\gamma} - c_{e})$ by a factor of -(1/2).}  

If $c_e<c_\gamma$ ($\delta < 0$), 
the decay of a photon into an electron-positron pair is kinematically allowed
for photons with energies exceeding

\begin{equation}
E_{\rm max}= m_e\,\sqrt{2/|\delta|}\;. 
\end{equation}

\noindent The  decay would take place rapidly, so that photons with energies 
exceeding $E_{\rm max}$ could not be observed either in the laboratory or as 
cosmic rays. Since photons have been observed with energies   
$E_{\gamma} \ge$ 50~TeV from the Crab nebula \cite{ta98}, we deduce for this 
case that $E_{\rm max}\ge 50\;$TeV, or that $|\delta| < 2\times  
10^{-16}$ \cite{sg01}. Stronger bounds on $\delta$ can be set
through observations of very high energy (TeV) photons.
The detection of cosmic $\gamma$-rays with energies
greater that 50~TeV from sources within our galaxy would improve the bound on
$\delta$.

If, on the other hand, $c_e>c_\gamma$ ($\delta > 0$) and electrons become 
superluminal if their energies exceed $E_{\rm max}/\sqrt{2}$.
Electrons traveling faster than light will emit light  at all frequencies by a
process of `vacuum \v{C}erenkov radiation.' This process occurs rapidly, so
that superluminal electron energies quickly approach $E_{\rm max}/\sqrt{2}$. 
Because electrons have been seen in the cosmic radiation 
with energies up to $\sim\,$2~TeV\cite{ni00}, it follows that 
$\delta <  3 \times 10^{-14}$. This upper limit is about two orders of
magnitude weaker than the limit obtained for Case I. 
  
A smaller, but more indirect, upper limit on $\delta$ for the $\delta > 0$
case can be obtained from theoretical considerations of \gray\ emission from 
the Crab Nebula. The unpulsed \gray\ spectrum of the Crab Nebula can be 
understood to be produced by sychrotron emission up to the $\sim 0.1$ GeV 
\gray\ energy range. 
Above 25 MeV, the synchrotron component falls off very rapidly with energy
as expected from theoretical limits on electron acceleration \cite{de96}. 
Emission above 0.1 GeV, extending into the TeV range, can be explained as
synchrotron self-Compton emission of the same relativistic electrons which
produce the synchrotron radiation. The Compton component, extending to 50
TeV, implies the existence of electrons having energies at least this great
in order to produce 50 TeV photons, even in the extreme Klein-Nishina limit.
This is, of course, required by conservation of energy.
This indirect argument, based on the reasonable assumption that the 50 TeV
\grays\ are from Compton interactions, leads to a smaller upper limit on 
$\delta$, {\it viz.,} $\delta < 10^{-16}$. Better observational \gray\ data 
for the Crab Nebula at multi-TeV energies, and data on the shape of its \gray\ 
spectrum above 50 TeV, will be needed to further test this indirect 
constraint on $\delta$. One open question involves the possibility of
hadronic interactions involving cosmic ray nucleons producing $\pi^0$'s 
which decay into observed high energy \grays\ (although there is no present 
indication of this). Hopefully, better data will determine whether or not
there is a significant hadronically induced component contributing to the 
multi-TeV spectrum of the Crab.

A further constraint on $\delta$ for $\delta > 0$ ($c_e>c_\gamma$)
follows from the modification of the threshold energy for the pair 
production process $\gamma + \gamma \rightarrow e^+ + e^-$. This follows from
the fact that the square of the four-momentum is changed to give the
threshold condition

\begin{equation}
2\epsilon E_{\gamma}(1 - \cos \theta) - 2E_{\gamma}^2\delta  >4 m_{e}^2,
\end{equation}

\noindent where $\epsilon$ is the energy of the low energy photon and $\theta$
is the angle between the two photons. The second term on the left-hand-side
comes from the fact that $c_{\gamma} =  
\partial E_{\gamma}/\partial p_{\gamma}$.

For head-on collisions ($\cos \theta = -1$) the minimum low energy photon
energy for pair production becomes 

\begin{equation}
\epsilon_{min} = {m_{e}^2\over{E_{\gamma}}} +  {E_{\gamma}\,\delta\over{2}}.
\end{equation}

It follows that the condition for a significant increase in the energy
threshold for pair production is $E_{\gamma}\delta/2$ $ \ge$
$ m_{e}^2/E_{\gamma}$, or 
equivalently, 

\begin{equation}
\delta \ge {2m_{e}^{2}\over{E_{\gamma}^{2}}}.
\end{equation}

As discussed in the previous section, there is no indication of LI violation
suppressing the physics of pair production for photons up to an energy of
$\sim20$ TeV. Thus, it follows from eq. (6) that the Mkn 501 observations
imply the constraint $\delta \le 2m_{e}^{2}/E_{\gamma}^{2} = 
1.3 \times 10^{-15}$ \cite{sg01}. This constraint on positive $\delta$ is 
more secure than the smaller, but indirect, limit given by the Crab Nebula 
acceleration model.

\section{Quantum Gravity Models}

In the absence of a true and complete theory of quantum gravity, theorists 
have been suggesting and exploring models to provide experimental and 
observational tests of possible manifestations of quantum gravity phenomena. 
Such phenomena have usually been suggested to be a possible result of quantum 
fluctuations on the Planck scale $\mpl = \sqrt{\hbar c/G} \simeq 1.22 
\times 10^{19}$ GeV/c$^2$, corresponding to a length scale $\sim 1.6 \times 
10^{-35}$ m \cite{ga95} - \cite{al02}. In models involving large extra 
dimensions, the energy scale at which gravity becomes strong can be much 
smaller than $\mpl$, with the quantum gravity scale, $\mqg$, approaching the 
TeV scale \cite{el00}, \cite{el01}.

In many of these models Lorentz invariance is predicted to be violated at high 
energy. This results in interesting modifications of particle physics that 
are accesible to observational tests using TeV \gray\ telescopes and cosmic 
ray detectors. An example of such a model is a quantum gravity model 
with a preferred inertial frame given by the cosmological rest frame of the 
cosmic microwave background radiation (For an extensive discussion, see the
review given in Ref. \cite{sm03}.) 

In the most commonly considered of these models, the usual relativistic 
dispersion relations between energy and momentum of the photon and the electron

\begin{equation}
E_{\gamma}^2 = p_{\gamma}^2 
\end{equation}

\begin{equation}
E_{e}^2 = p_{e}^2 + m_{e}^2
\end{equation}

(with the ``low energy'' speed of light, $c \equiv 1$) are modified by a 
leading order quantum space-time geometry corrections which
are cubic in $p \simeq E$ and are suppressed by the quantum gravity mass 
scale $M_{QG}$. Following Refs. \cite{am98} and \cite{al02}, we take the 
modified dispersion relations to be of the form

\begin{equation}
E_{\gamma}^2~ =~ p_{\gamma}^2~ -~ {p_{\gamma}{^3}\over M_{QG}} 
\end{equation}

\begin{equation}
E_{e}^2~ =~ p_{e}^2~ +~ m_{e}^2 ~-~ {p_{e}{^3}\over M_{QG}} 
\end{equation}

We assume that the cubic 
terms are the same for the photon and electron as in eqs. (9) and (10). More 
general formulations have been considered by Jacobson, Liberati and 
Mattingly \cite{jlm02} and Konopka and Major \cite{ko02}. 

We note that there are variants of quantum gravity and large extra dimension
models which do not violate Lorentz invariance and for which the constraints
considered here do not apply. There are also variants for which there
are no cubic terms in momentum, but rather terms of the quartic form 
$\sim {p{^4}/ M_{QG}^2}$. These terms are suppressed by two orders
of the Planck mass and are therefore much smaller and do not violate
the constraints derived here. 

As opposed to the Coleman-Glashow formalism \cite{co99}, which involves mass 
dimension four
operators in the Lagrangian and preserves power-counting renormalizablility,
the cubic term which modifies the dispersion relations may be considered in 
the context of an effective 
``low energy'' field theory, valid for $E \ll M_{QG}$, in which case the 
cubic term is a small perturbation involving dimension five operators 
whose construction is discussed in Ref. \cite{my03}. With this {\it caveat}, 
we can 
generalize the LI violation parameter $\delta$ to an energy dependent form

\begin{equation}
\delta~ \equiv~ {\partial E_{e}\over{\partial p_{e}}}~ -~ {\partial E_{\gamma}
\over{\partial p_{\gamma}}}~
 \simeq~ {E_{\gamma}\over{M_{QG}}}~ 
-~{m_{e}^{2}\over{2E_{e}^{2}}}~ -~ {E_{e}\over{M_{QG}}} ,
\end{equation}

which is a valid approximation for the energy regime $E_{e} \gg m_{e}$.
Note that the maximum velocities of particles of type $i$ are reduced
by ${\cal{O}}(E_{i}/M_{QG})$.

For pair production then, with the positron and electron energy 
$E_{e} \simeq E_{\gamma}/2$,

\begin{equation}
\delta~ =~ {E_{\gamma}\over{2M_{QG}}}~ -~ {2m_{e}^{2}\over{E_{\gamma}^{2}}} 
\end{equation}

and the threshold condition given by eq.(6) reduces to the constraint

\begin{equation}
M_{QG} ~\ge~ {E_{\gamma}^3\over 8m_{e}^2}.
\end{equation}

Since pair production occurs for energies of at least 20 TeV, as indicated
by our analyses of the Mkn 501 and Mkn 421 spectra \cite{ds02},\cite{ko03}, 
we then find the constraint on the quantum gravity scale
$M_{QG} \ge 0.3 M_{Planck}$. This constriant contradicts the 
predictions of some proposed quantum gravity models involving large extra 
dimensions and smaller effective Planck masses. Previous constraints on 
$M_{\rm QG}$ for the cubic model, obtained from limits on the energy 
dependent velocity dispersion of \grays\ for a TeV flare in Mkn 421 
\cite{bi99} and from \gray\ bursts \cite{sc99} were in the much less 
restrictive range $M_{QG} \ge (5-7) \times 10^{-3} M_{Planck}$.

Within the context of a more general cubic modification of the dispersion 
relations given by eqs. (9) and (10), Jacobson, {\it et al.} \cite{jlm03} 
have obtained an indirect limit on $M_{QG}$ from the apparent cutoff in the 
synchrotron component of the in the Crab Nebula \gray\ emission at $\sim 0.1$ 
GeV. By making reasonable assumptions to modify the standard synchrotron
radiation formula to allow for Lorentz invariance violation, they have 
concluded that the maximium synchrotron photon energy will be given by 
$E_{\gamma, {\rm max}}$ = 0.34 $(eB/m_{e})(m_{e}/M_{QG})^{-2/3}.$
This reasoning leads to the constraint $ M_{QG}$$ >$$ 1.2 \times 10^{7} 
M_{Planck}.$

Future observations of the Crab Nebula with the {\it GLAST} (Gamma-Ray 
Large Area Space Telescope) satellite, scheduled to be launched in 2005,
will provide a better determination of its unpulsed \gray\ spectrum 
in the energy range above 30 MeV where the transition from the synchrotron
emission component to the Compton emission component occurs. This will
provide a more precise determination of the maximum electron energy in the
Nebula and therefore provide a more precise constraint on the parameter
$M_{QG}$ as we have defined it here. However,
this constraint will still be orders of magnitude above the Planck scale.

\section{Conclusions}

Nearly a century after the inception of special relativity, high energy 
\gray\ observations have confirmed its validity up to electron energies
of 2 TeV, photon energies of 20 TeV and, indirectly, up to electron energies
in the PeV range. These results indicate an absence of evidence for 
proposed violations of Lorentz invariance as predicted by some 
phenomenological quantum gravity and large extra dimension models. Thus, high 
energy astrophysics has provided important empirical constraints on Planck 
scale physics.

Some models with large extra dimensions \cite{el00}, \cite{el01} are ruled out 
by the existence of absorption in the very high energy 
spectra of nearby BL Lac objects. The fact that more distant 
brighter sources are not seen can also be taken as indirect evidence of 
intergalactic absorption by pair production interactions \cite{st96}.
The constraints based on analysis of the Crab Nebula \gray\ spectrum,
discussed in the previous section, imply that the quantum
gravity scale is orders of magnitude above the Planck mass scale. This
indicates that the class of models considered here with linear Planck
scale suppressed terms in the dispersion relations cannot be
reflective of physics at the Planck scale. Models such as loop quantum gravity 
with a preferred inertial frame are ruled out by this line of reasoning. 
Alternative models to consider might be models with  
quartic momentum terms with $M_{QG}^2$ supression in the dispersion relations, 
Lorentz invariant quantum gravity models, or really new Planck scale 
physics such as string theory, which preserves Lorentz invariance.

\section*{Acknowledgments}
I would like to thank Ted Jacobson, Stefano Liberati, David Mattingly and
Serge Rudaz for helpful discussions.

\end{document}